\documentclass[conference]{IEEEtran}
\IEEEoverridecommandlockouts
\usepackage{cite}
\usepackage{amsmath,amssymb,amsfonts}
\usepackage{algorithmic}
\usepackage{graphicx}
\usepackage{textcomp}
\usepackage{xcolor}
\usepackage{upgreek}
\usepackage{mathtools}
\DeclareMathOperator*{\argmin}{arg\,min}

\DeclarePairedDelimiter\set\{\}
\def\BibTeX{{\rm B\kern-.05em{\sc i\kern-.025em b}\kern-.08em
    T\kern-.1667em\lower.7ex\hbox{E}\kern-.125emX}}
\begin{document}

\title{Accurate Real-Time Estimation of 2-Dimensional Direction of Arrival using a 3-Microphone Array \\
\thanks{This work was supported by the National Institute on Deafness and Other Communication Disorders (NIDCD) of the National Institutes of Health (NIH) under Award 5R01DC015430-05. The content is solely the responsibility of the authors and does not necessarily represent the official views of the NIH.}
}

\author{\IEEEauthorblockN{1\textsuperscript{st} Anton Kovalyov}
\IEEEauthorblockA{\textit{Electrical and Computer Engineering} \\
\textit{University of Texas at Dallas}\\
Richardson, USA \\
anton.kovalyov@utdallas.edu}
\and
\IEEEauthorblockN{2\textsuperscript{nd} Kashyap Patel}
\IEEEauthorblockA{\textit{Electrical and Computer Engineering} \\
\textit{University of Texas at Dallas}\\
Richardson, USA \\
patelkashyap@utdallas.edu}
\and
\IEEEauthorblockN{3\textsuperscript{rd} Issa Panahi}
\IEEEauthorblockA{\textit{Electrical and Computer Engineering} \\ 
\textit{University of Texas at Dallas}\\
Richardson, USA \\
imp015000@utdallas.edu}
}

\maketitle

\begin{abstract}
This paper presents a method for real-time estimation of 2-dimensional direction of arrival (2D-DOA) of one or more sound sources using a nonlinear 3-microphone array. 2D-DOA is estimated employing frame-level time difference of arrival (TDOA) measurements. Unlike conventional methods, which infer location parameters from TDOAs using a theoretical model, we propose a more practical approach based on supervised learning. The proposed model employs nearest neighbor search (NNS) applied to a spherical Fibonacci lattice consisting of TDOA to 2D-DOA mappings learned in the field. Filtering and clustering post-processors are also introduced for improving source detection and localization robustness.\par
\end{abstract}

\begin{IEEEkeywords}
DOA, TDOA, array, nearest neighbor, clustering
\end{IEEEkeywords}

\section{Introduction}
Time difference of arrival (TDOA)-based sound source localization (SSL) is a well-established approach in the literature. When a source is at far-field from the array, or the number of microphones is less than four, 3-dimensional (3D) TDOA-based SSL is not possible, and direction of arrival (DOA) is estimated instead. Practical applications include beamforming \cite{beamforming_doa}, blind source separation \cite{bss_doa}, and to provide a visual DOA indicator for people with spatial hearing loss (SHL) \cite{doa_2_mic_app1, doa_3_mic_app1, doa_3_mic_app2}. In this work, we are especially interested in the latter.\par

Anshuman et al. \cite{doa_2_mic_app1} proposed a smartphone application for providing a visual azimuthal DOA indicator of a speech source for people with SHL. The use of a smartphone for this purpose is especially convenient due to its widespread availability. However, DOA is estimated using only two microphones, resulting in what is known as the \textit{front-back} ambiguity, which is a common issue in linear arrays. Nowadays many smartphones have an array of at least three microphones. When the array is nonlinear, not only do we avoid front-back ambiguity, but we also allow estimating both azimuth and elevation angles of a source, known as 2D-DOA estimation. Tokg\"{o}z et al. \cite{doa_3_mic_app1, doa_3_mic_app2} proposed adaptations of the work in \cite{doa_2_mic_app1} for \textit{L}-shaped arrays of three microphones. However, these methods are constrained to the detection and localization of an individual speech source. Additionally, no practical scheme is proposed to compensate for discrepancies in measured and theoretical TDOAs, which are common in practical systems for reasons such as erroneous array calibration and varying phase response of microphones.\par

\begin{figure}[htb!]
    \centering
    \includegraphics[trim={0 0 0 0},clip, width=0.40\textwidth]{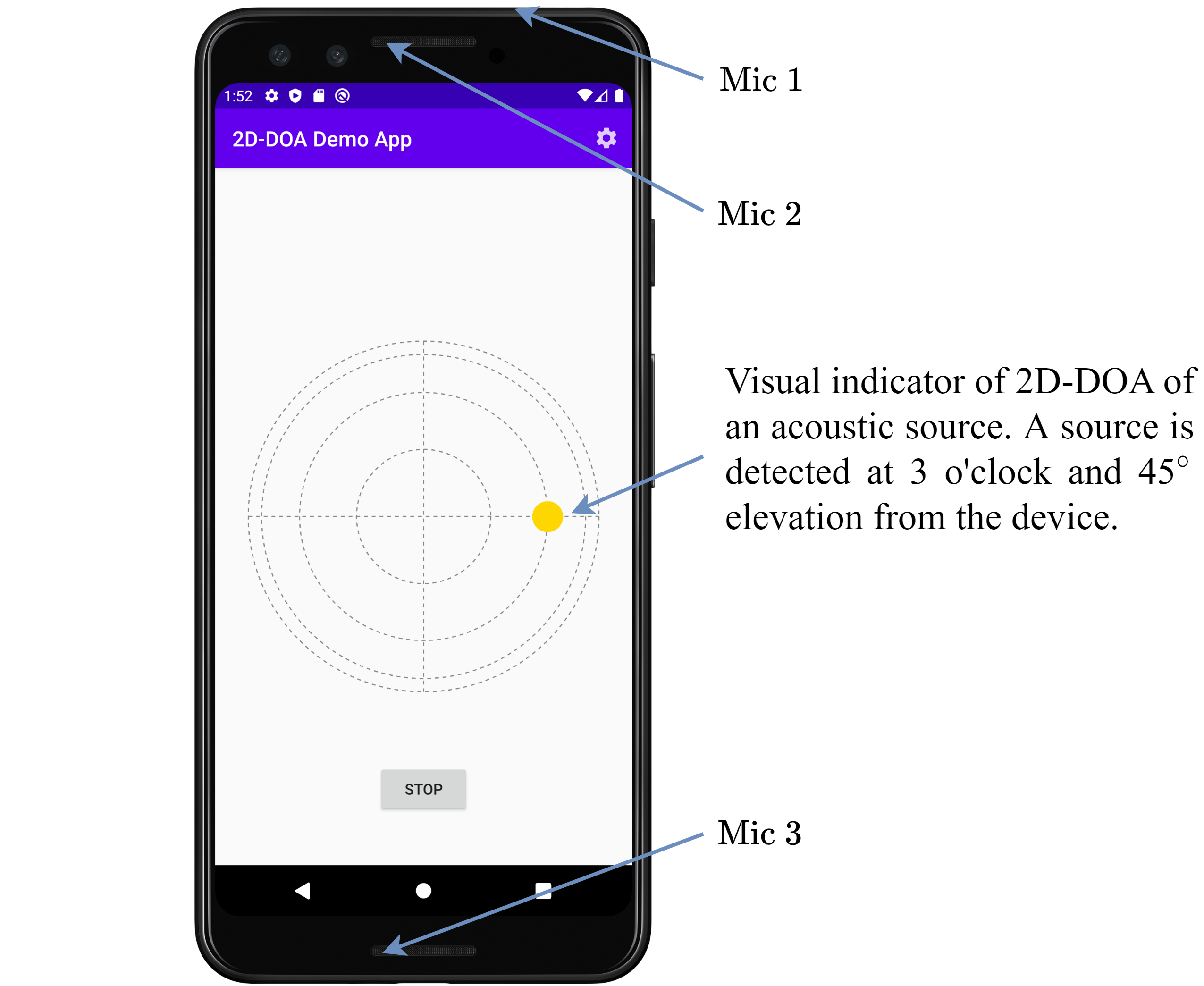}
    \caption{The proposed method deployed on a smartphone application to provide a visual 2D-DOA indicator (orthographic projection of the source position on a hemisphere above the device) for people with SHL.}
    \label{fig:demo_app_doa}
\end{figure}

Motivated by the above observations, we propose a practical method for real-time 2D-DOA estimation of multiple sound sources using a nonlinear 3-microphone array. Instead of a theoretical model, TDOA to 2D-DOA mappings are learned in the field and inference is performed applying nearest neighbor search (NNS) to a spherical Fibonacci lattice containing the learned mappings. Furthermore, filtering and clustering post-processors, designed for reliable detection and localization of one or more sources, are also introduced. As shown in Fig. \ref{fig:demo_app_doa}, the proposed method was implemented on a smartphone.\par

\section{Problem Formulation} \label{sec:problem}
Let us consider an array of three microphones in a noisy and reverberant environment with $C$ acoustic sources. Let $\mathbf{m}_i$ and $\mathbf{s}_c$ denote the $i$-th microphone and $c$-th source 3D positions, for $i \in \set{1,2,3}$ and $c \in \set{1,2,\ldots,C}$. The signal captured by the $i$-th microphone is modeled by
\begin{equation}
    \mathbf{y}_i = \mathbf{v}_i 
 + \sum_{c=1}^C \mathbf{x}_{ic} \;,
\end{equation}
where $\mathbf{v}_i$ is incoherent noise and $\mathbf{x}_{ic}$ is the reverberant signal of the $c$-th source. Assuming sufficient angular spacing, the problem is formulated as real-time detection of the $C$ sources and estimation of their azimuth and elevation angles with respect to the microphone array.\par

\section{Methodology} \label{sec:method}
Let us segment all $\mathbf{y}_i$ into $K$ overlapping frames of length $L$. The proposed method operates at frame-level in a causal manner. The processing pipeline consists of four stages: (1) TDOA estimation; (2) mapping measured TDOAs to 2D-DOA; (3) filtering unreliable estimates; and (4) clustering.\par

\subsection{TDOA Estimation} \label{sec:tdoa}
Let $\mathbf{y}_{ik}$ be the $k$-th segment of $\mathbf{y}_i$, for $k = 1,2,\ldots,K$. Let $V$ be the subset of frame indices for which exactly one dominant source is present at a time. It is assumed that $V$ is not empty and it includes frame indices corresponding to every source in the mixture, which, as long as the segment length is not made too long, are reasonable assumptions, especially for sparse signals such as speech. Let $\mathbf{s}(v)$, denote the 3D position of the dominant source at frame index $v \in V$. The TDOA in meters of the direct path signal originating at $\mathbf{s}(v)$ when received between $\mathbf{m}_i$ and $\mathbf{m}_j$, for $j \in \set{1,2,3}$ such that $j \neq i$, is given by
\begin{equation} \label{eq:tdoa}
    r_{ij}(v) = ||\mathbf{s}(v) - \mathbf{m}_i|| - ||\mathbf{s}(v) - \mathbf{m}_j|| \;.
\end{equation}
Assuming the signal propagation speed is known, an estimate of $r_{ij}(v)$ can be found by the peak of some weighted cross-correlation function between $\mathbf{y}_{iv}$ and $\mathbf{y}_{jv}$ \cite{TDOA_GCC}. Here we use modified Cross-Power Spectrum Phase \cite{TDOA_modified_PHAT} (mCPSP) paired with quadratic interpolation \cite{TDOA_QI} (QI) for improved estimation resolution. TDOAs are estimated for every $k$-th frame. Unreliable estimates are rejected at the filtering stage.\par



\subsection{Accurate mapping of TDOAs to 2D-DOA} \label{sec:mappings}
\begin{figure}[htb!]
    \centering
    \includegraphics[trim={0 0 0 0},clip, width=0.48\textwidth]{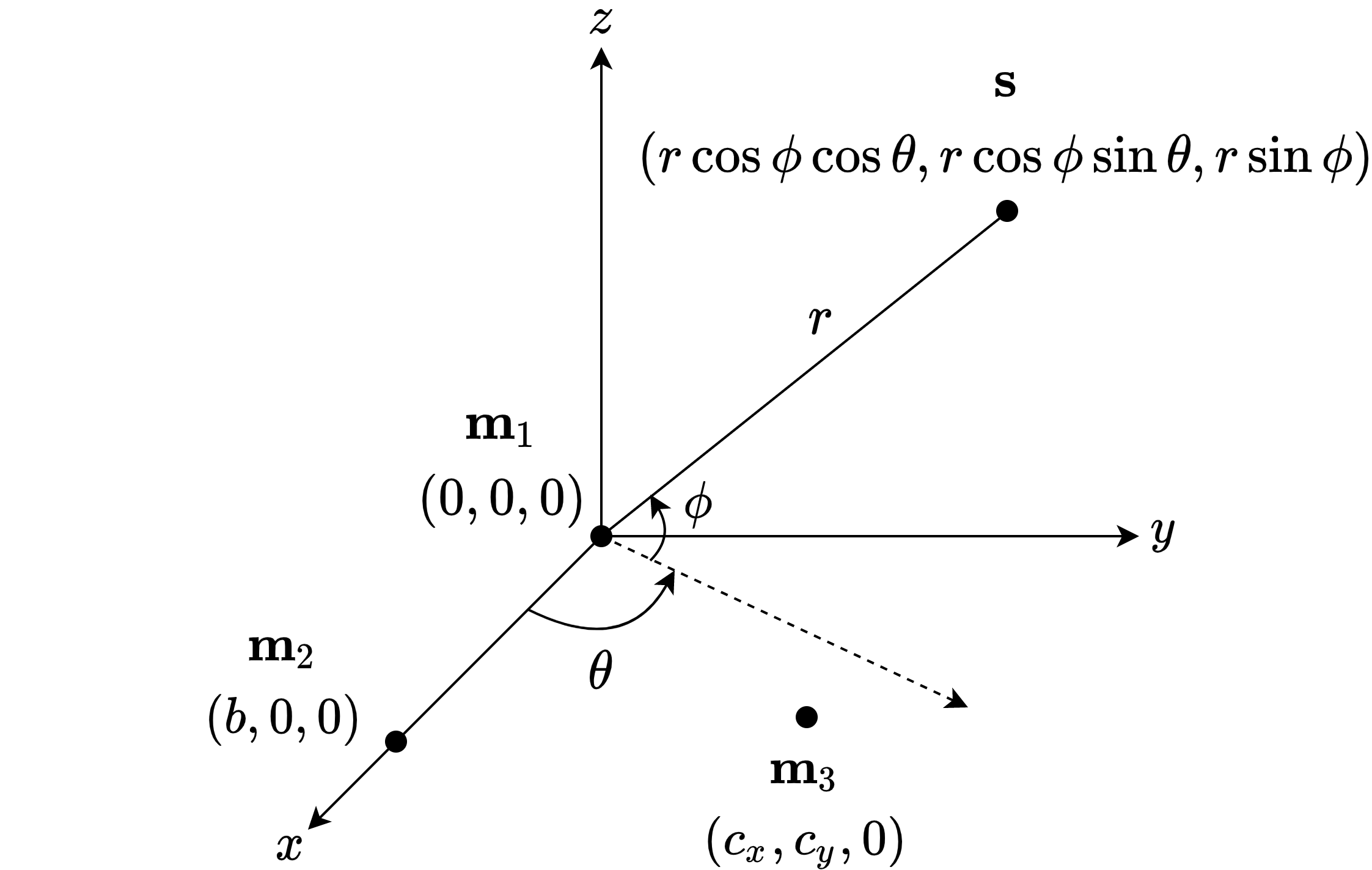} \vspace{0.0\textwidth}
    \caption{Problem geometry of 2D-DOA estimation with three microphones.}
    \label{fig:cf_doa}
\end{figure}

Following the formulation of the previous section, we consider frame-level localization of a single source. Hence, for simplicity, we drop the source and frame indices notation unless an ambiguity arises. Let $\mathbf{s}$, $r_{ij}$, and $\hat{r}_{ij}$ denote the 3D source position, the true TDOA in meters and its corresponding estimate, respectively. To have better insight into the problem geometry let us first consider a simple closed form solution (CF) that maps TDOAs to 2D-DOA. To reduce degrees of freedom, we fix $\mathbf{m}_1$ at the origin, i.e., $[0,0,0]^T$, $\mathbf{m}_2$ at $[b,0,0]^T$, and $\mathbf{m}_3$ at $[c_x,c_y,0]^T$. Let $r$ be the distance from source to origin. When $r$ is large, i.e., the source is at far field, its actual value has negligible impact on TDOA. Thus, we reduce another degree of freedom by fixing $r$ to some large value. Fig. \ref{fig:cf_doa} illustrates the problem geometry. The parameters of interest are $\theta$ and $\phi$, which are the azimuth and elevation angles of the source, respectively. For simplicity, let us reparametrize the problem into finding the source 3D position $\mathbf{s}$ on a sphere of far-field radius $r$ centered at the origin, where $\theta$ and $\phi$ can be inferred from $\mathbf{s}$ using the relationship in Fig. \ref{fig:cf_doa}. Let $s_x$, $s_y$, and $s_z$ represent the corresponding $xyz$ coordinates of $\mathbf{s}$. A CF mapping linearly independent TDOAs $r_{12}$ and $r_{13}$ to $\mathbf{s}$ is given by
\begin{equation} \label{eq:cf_doa}
    \begin{split}
        s_x &= \dfrac{b^2 + 2r_{12}r - r_{12}^2}{2b} \\
        s_y &= \dfrac{c_x^2 + c_y^2 - r_{13}^2 + 2r_{13}r - 2c_x s_x}{2c_y} \\
        s_z &= \pm \sqrt{r^2 - s_x^2 - s_y^2}\;.       
    \end{split}
\end{equation}
We notice that $b$ and $c_y$ cannot be 0, meaning that a nonlinear array is needed to estimate 2D-DOA. We further note that $s_z$ can either be negative or positive. Throughout this work we let $s_z$ be positive. The problem can then be visualized as localizing a source on a hemisphere above the array, which is equivalent to letting $\theta \in \left[-\pi, \pi\right]$ and $\phi \in \left[0, \pi/2 \right]$.\par

If measured TDOAs do not match theoretical values for reasons other than reverberation and noise, e.g., erroneous array calibration and varying phase response of microphones, the CF in (\ref{eq:cf_doa}) may not be accurate. For improved performance, TDOA to 2D-DOA mappings can be learned directly in the field. Consequently, we propose the following supervised learning approach. Let us discretize the search space into
\begin{equation}
    S = \set{\boldsymbol\upgamma^{(1)}, \boldsymbol\upgamma^{(2)}, \ldots, \boldsymbol\upgamma^{(N)}} \;,
\end{equation}
where $\boldsymbol\upgamma^{(n)} = [\theta^{(n)}, \phi^{(n)}]^T$, for $n = 1,2,...,N$, groups the $n$-th 2D-DOA tuple in the search space. Similarly, let 
\begin{equation} \label{eq:tdoa_mappings}
    Q = \set{\mathbf{q}^{(1)}, \mathbf{q}^{(2)}, \ldots, \mathbf{q}^{(N)}} \;,
\end{equation}
where $\mathbf{q}^{(n)} = [r_{12}^{(n)}, r_{13}^{(n)}, r_{23}^{(n)}]^T$ groups corresponding TDOA mappings. All combinations are considered for robustness. The TDOA mappings in $Q$ are collected in a supervised manner offline. During inference, NNS is applied as follows
\begin{equation} \label{eq:nns_estimator}
    \boldsymbol{\hat{\upgamma}} = \argmin_{\boldsymbol\upgamma^{(n)}}||\mathbf{q}^{(n)} - \mathbf{\hat{q}}||^2 \;,
\end{equation}
where $\mathbf{\hat{q}} = \left[\hat{r}_{12}, \hat{r}_{13}, \hat{r}_{23}\right]^T$. Mappings are stored in a $k$-d tree structure \cite{kd_tree_original, kd_tree_search}, allowing NNS in expected logarithmic time.\par


\begin{figure}[htb!]
    \centering
    \begin{tabular}{ c c }
        \includegraphics[trim={2.5cm 1cm 2.5cm 0.5cm},clip,width=0.2\textwidth]{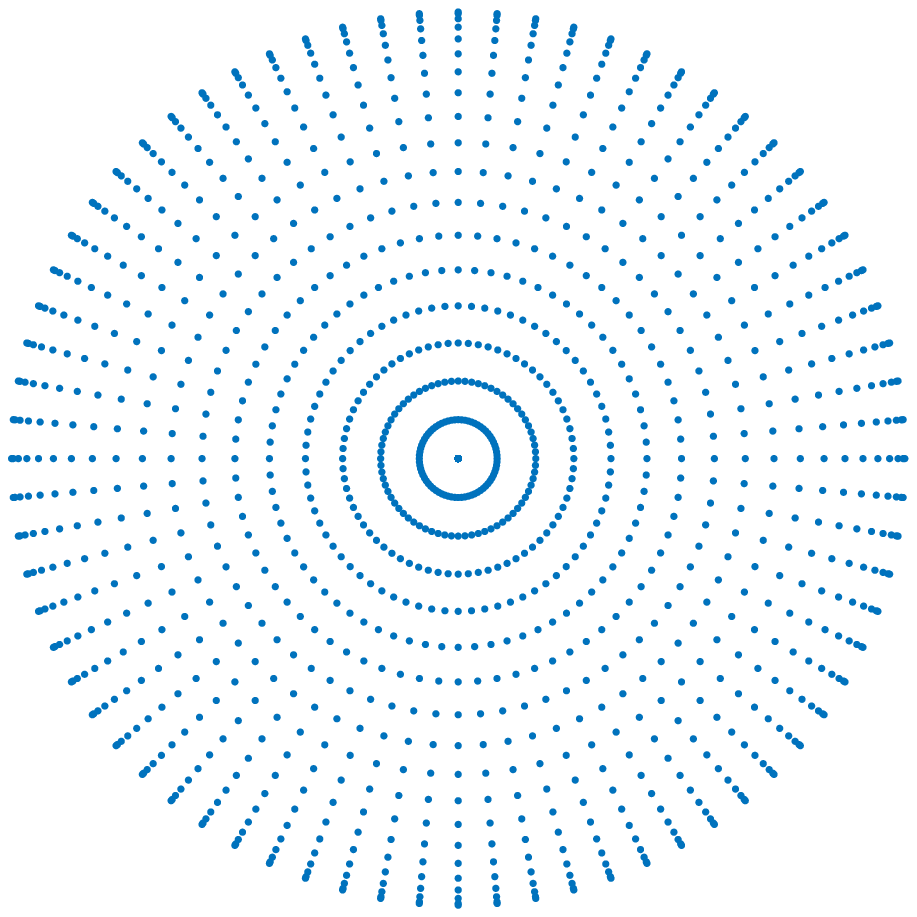} &         \includegraphics[trim={2.5cm 1cm 2.5cm 0.5cm},clip,width=0.2\textwidth]{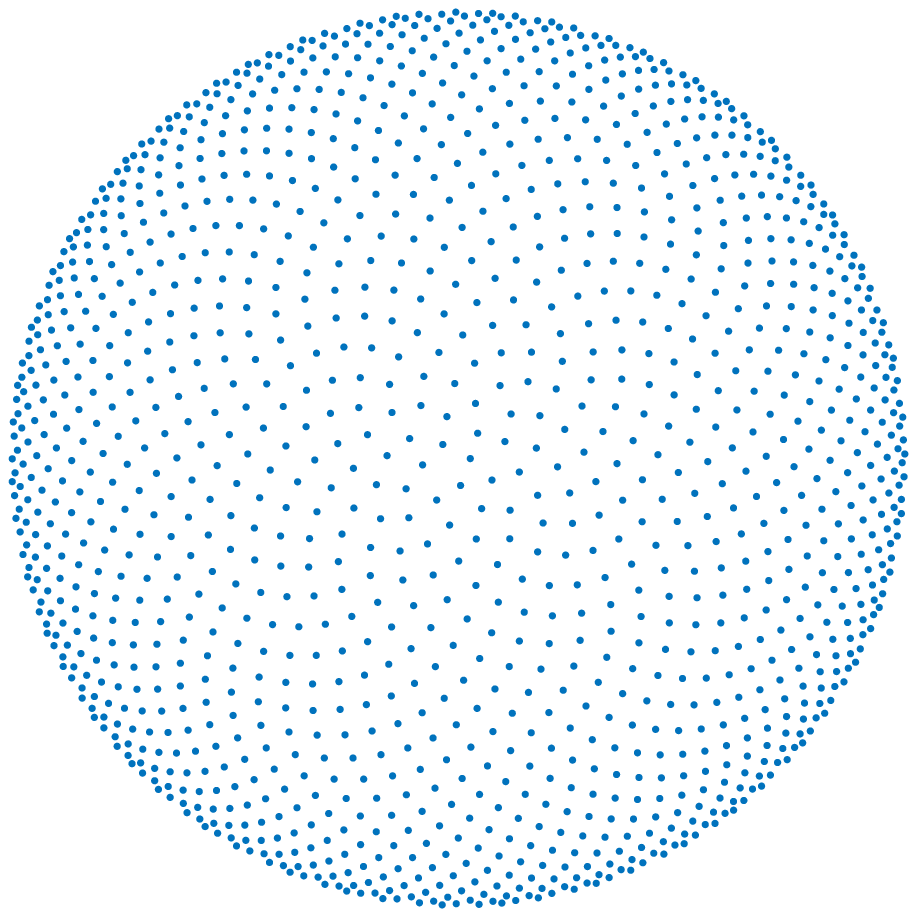} \\
        \hspace{0.0cm} \small (a) & \hspace{0.0cm} \small (b) \\
    \end{tabular}
    \caption{Two different methods for placing $N = 1297$ points on a hemisphere. Latitude-longitude lattice (a) and Fibonacci lattice (b). Orthographic projections centered at the pole.}
    \label{fig:lattice}
\end{figure}

For further efficiency, the candidate solutions in $S$ should be distributed as evenly as possible throughout a hemisphere. A simple approach is to discretize $\theta$ and $\phi$ by some angular spacing $\delta = 180^{\circ} / u$, where $u$ is a positive integer, resulting in a point distribution known as the \textit{latitude-longitude} lattice \cite{fibonacci_latitude_longitude}. This lattice can be visualized as a set of points placed at the intersections of a grid of meridians and parallels (Fig. \ref{fig:lattice}a). The total number of points is $N = u^2 + 1$, which comes from the number of meridians ($2u$) times the number of parallels ($u/2$) plus one pole. In this lattice, however, points concentrate around the pole, resulting in noticeable anisotropy. Instead, we apply the spherical Fibonacci point set algorithm \cite{fibonacci_hemisphere} for a more uniform point distribution (see Fig. \ref{fig:lattice}b). This lattice is generated by
\begin{equation} \label{eq:fib_lattice}
    \begin{split}
        \theta^{(n)} &= \dfrac{2\pi (n-1)}{\Phi} \\
        \phi^{(n)} &= \dfrac{\pi}{2} - \cos^{-1}{\bigg(1 - \dfrac{2n - 1}{2N}\bigg)} \;,
    \end{split}
\end{equation}
where $\Phi = (1 + \sqrt{5}) / 2$ is the golden ratio.\par 


\subsection{Filtering} \label{sec:filtering}
We propose a simple three-step filter that rejects unreliable sets of TDOA measurements, i.e., $\hat{\mathbf{q}}$ in (\ref{eq:nns_estimator}). The purpose is to filter out all measurements computed at frame index $k \not\in V$. Each sequential step consists of verifying that a measurement satisfies a certain reliability condition. The first two conditions are applied on individual TDOAs only. If any of the estimates does not satisfy a given condition, the entire set is rejected. The first condition verifies that there is acoustic activity at the estimated lag by ensuring that the cross-correlation value of the peak is above a positive threshold $T_R$, as given by
\begin{equation} \label{eq:filter1}
    R(\ell_{\text{max}}) > T_R \;,
\end{equation}
where $R(\ell)$ is the weighted cross-correlation function given by mCPSP at lag index $\ell$, and $\ell_{\text{max}}$ is the lag of the peak. In the second step, we quantify how dominant the cross-correlation peak is compared to other solution candidates by computing a confidence level $\beta \in [0,1]$, given by 
\begin{equation}
    \begin{split}
        \beta &= 1 - \dfrac{\eta}{R(\ell_{\text{max}})} \\
        \eta &= \dfrac{1}{|\mathcal{L}|} \sum_{\ell \in \mathcal{L}} \max\left\{0, {R(\ell)}\right\} \;,
    \end{split}
\end{equation}
where $\mathcal{L}$ is the set of all plausible lags not including $\ell_\text{max}$. We then verify that
\begin{equation}
    \beta > T_\beta \;,
\end{equation}
where $T_\beta$ is some threshold. Finally, the third step ensures coherence among TDOAs in $\hat{\mathbf{q}}$ by verifying that the error of the NNS estimator in (\ref{eq:nns_estimator}) is below a threshold $T_q$, as given by
\begin{equation}
    ||\mathbf{q}(\hat{\boldsymbol{\upgamma}}) - \hat{\mathbf{q}}||^2 < T_q \;,
\end{equation}
where $\mathbf{q}(\hat{\boldsymbol{\upgamma}})$ is the closest match to $\hat{\mathbf{q}}$ found by NNS.\par

\subsection{Clustering} \label{sec:clustering}
Clustering is used to assign frame-level 2D-DOA estimates to the correct source and combine accumulated measurements in such a way to improve source detection and localization reliability. For this purpose, we propose Recency and Frequency aware Exponential Filter Clustering (RFEFC). RFEFC is closely based on the exponential filtering concept used in Real-Time Exponential Filter Clustering (RTEFC) \cite{rtefc}. In RTEFC, exponential filtering is employed to update the cluster within minimum distance from a given location measurement. Apart from allowing real-time frame-level processing and low memory and computational overhead, RTEFC also exhibits good tracking capabilities. However, RTEFC does not take into account recency and frequency of data, which are important characteristics in the context of this work.\par

In RFEFC, a fixed number of clusters $N_c$ is maintained in real-time. Each $i$-th cluster, for $i \in \left\{1,2,\ldots,N_c \right\}$, consists of a cluster centroid $\hat{\mathbf{s}}_i$ and a cluster confidence level $\rho_i$. Here, $\hat{\mathbf{s}}_i$ represents an estimate of the $i$-th 3D source position on a unit radius hemisphere. The use of Cartesian representation is necessary for averaging. $\rho_i$, on the other hand, is a value between $0$ and $1$ quantifying the confidence of estimate $\hat{\mathbf{s}}_i$, with $1$ meaning high confidence. Initially, $\hat{\mathbf{s}}_i = \varnothing$ and $\rho_i = 0$ for all $i$, meaning the clusters are inactive, i.e., no source is detected. Let $\hat{\mathbf{s}}$ be a frame-level estimate of some source location sampled every $\Delta t$ seconds. If the estimate was rejected by the filtering process, we let $\hat{\mathbf{s}} = \varnothing$. RFEFC searches for the cluster with maximum confidence level whose centroid is within some minimum distance $d_{\text{min}}$ from $\hat{\mathbf{s}}$. If such cluster is found, its centroid $\hat{\mathbf{s}}_{\text{close}}$ is updated using exponential filtering and its confidence level $\rho_{\text{close}}$ is increased as follows
\begin{equation} \label{eq:exp_filtering}
    \begin{split}
        \hat{\mathbf{s}}_{\text{close}} &= \alpha \mathbf{\hat{s}}_{\text{close}} + (1-\alpha) \hat{\mathbf{s}} \\
        \rho_{\text{close}} &= \min{\left\{1, \rho_{\text{close}} + N_s^{-1}\right\}} \;,
    \end{split}    
\end{equation}
where $N_s$ is some positive integer representing the number of consecutive frames needed for $\rho_\text{close}$ to reach $1$, thus keeping track of the \textit{frequency} with which a cluster is updated. If a cluster whose centroid is within $d_\text{min}$ distance from $\hat{\mathbf{s}}$ is not found, RFEFC selects the cluster with lowest confidence level and updates it as
\begin{equation}
    \begin{split}
        \hat{\mathbf{s}}_{\text{old}} &= \hat{\mathbf{s}} \\
        \rho_{\text{old}} &= N_s^{-1} \;.
    \end{split}
\end{equation}
This update can be interpreted as creating a new cluster from stale data. Finally, to give a sense of \textit{recency}, for every $i$-th cluster that was not updated, RFEFC decreases the confidence threshold followed by forgetting clusters whose confidence level reaches $0$. This sequence of operations is given by
\begin{equation}
    \begin{split}
        \rho_i &= \max{\left\{0, \rho_i - \frac{\Delta t}{T_\text{win}} \right\}} \\
        \hat{\mathbf{s}}_i &= 
        \begin{cases}
            \hat{\mathbf{s}}_i & \text{if } \rho_i > 0 \\
            \varnothing & \text{if } \rho_i = 0 \;,
        \end{cases}
    \end{split}    
\end{equation}
where $T_{\text{win}}$ is some positive real number controlling the time in seconds it takes RFEFC to forget an inactive cluster.\par 

The introduction of $\rho_i$ in RFEFC allows a simple mechanism to decide if a source is present at an estimated location. Here, a source $i$ is detected once $\rho_i$ reaches $1$ and remains above some threshold $T_a$. The memory and computational complexities of RFEFC are linear in $N_c$, thus making RFEFC similarly efficient to RTEFC. Also, due to exponential filtering, RFEFC can in principle track a moving source that produces continuous sound. Finally, unlike RTEFC, which updates the cluster closest to the location estimate, RFEFC updates the cluster with highest confidence level within a specified distance. The purpose of this selection rule is to minimize the likelihood of multiple closely spaced clusters being active for prolonged periods of time due to jittery estimates. In such scenarios, RFEFC would only update the cluster with highest confidence, thus forcing other nearby clusters to be forgotten.\par


\section{Experiments} \label{sec:experiments}
Three experiments were conducted in the field to evaluate the performance of the proposed method using the nonlinear 3-microphone array of a Pixel 3 smartphone, shown in Fig. \ref{fig:demo_app_doa}. Experiments were conducted in a moderately sized office room with typical office noise.\par 

The data collection, needed to populate $Q$ in (\ref{eq:tdoa_mappings}), was conducted as follows. The smartphone was placed on a turntable and a speaker was placed on a fixed surface at a distance of 0.6 meters from the turntable and at ten varying elevations ranging uniformly from $0^\circ$ to $90^\circ$. During data collection, the speaker played white noise while the turntable turned at a fixed rate till it made a full revolution, thus covering the entire azimuth range. White noise was the source signal due to its good auto-correlation properties. TDOAs were measured and mapped to known points on a latitude-longitude lattice. Interpolation was applied to generate the Fibonacci lattice in (\ref{eq:fib_lattice}).\par

The parameters were defined as follows. The sampling rate $f_s$ was 48 kHz and a frame length $L = 1024$ with 50\% overlap was used. The number of lattice points $N$ in Section \ref{sec:mappings} was set to $10^4$. The thresholds $T_R$, $T_\beta$, and $T_q$ in Section \ref{sec:filtering} were set to 1e-2, 0.5, and 5e-5, respectively. The parameters $N_c$, $\Delta t$, $d_{\text{min}}$, $N_s$, $\alpha$, $T_\text{win}$, and $T_a$ in Section \ref{sec:clustering} were set to $10$, $L/(2 f_s)$ s, 0.25, 5, 0.75, 5 s, and 0.5, respectively.\par

\begin{figure}[htb!]
    \centering
    \includegraphics[trim={0 0 0 0},clip, width=0.38\textwidth]{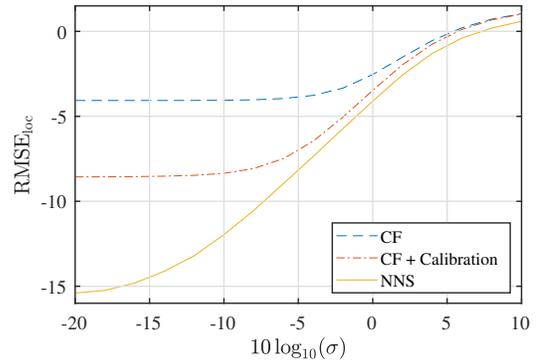}
    \caption{Experiment 1. 2D-DOA estimation performance of NNS vs. CF for varying noise in TDOA measurements.}
    \label{fig:rmse_loc_doa}
\end{figure}

In the first experiment, we use collected TDOAs to assess the practical effectiveness of supervised learning vs. simpler CF in mapping TDOAs to 2D-DOA, with the latter being the type of mapping scheme used in \cite{doa_2_mic_app1, doa_3_mic_app1,doa_3_mic_app2}. For this purpose, we benchmark 2D-DOA estimation performance of NNS in (\ref{eq:nns_estimator}) against its CF counterpart in (\ref{eq:cf_doa}). Localization root mean square error ($\text{RMSE}_{\text{loc}}$) is used as the performance metric. $\text{RMSE}_{\text{loc}}$ measures the localization error on a unit hemisphere over $M$ trials, as given by
\begin{equation}
    \text{RMSE}_\text{loc} = \sqrt{\dfrac{1}{M} \sum_{i=1}^M ||\mathbf{s}^{(i)} - \hat{\mathbf{s}}^{(i)}||^2} \;,
\end{equation}
where $\mathbf{s}^{(i)}$ is the source position on a unit hemisphere generated during the $i$-th trial, and $\hat{\mathbf{s}}^{(i)}$ is the corresponding estimate. To generate $\mathbf{s}^{(i)}$ for evaluation purposes, azimuth and elevation angles were drawn independently and uniformly at random from their respective ranges and mapped to Cartesian coordinates. Corresponding TDOAs were interpolated using the collected dataset in the field. Finally, to simulate measurement noise, interpolated TDOAs were corrupted using additive white Gaussian noise (AWGN) with varying standard deviation $\sigma \in \left[0.01, 10\right]$ cm. Since CF requires accurate knowledge of microphone array geometry, we consider two variants. In the first variant, simply referred to as CF, the array geometry is measured by hand. In the second variant (CF + Calibration), we use the dataset collected for NNS to calibrate the array applying Levenberg–Marquardt as the optimization algorithm. Fig. \ref{fig:rmse_loc_doa} reports the results. We note that CF without calibration exhibits somewhat poor performance, which is attributed to its sensitivity to non-precise microphone position estimates. Although introducing calibration improves results considerably, this scheme is still not sufficient to attain the high performance exhibited by NNS, especially at low and moderate noise. As a result, NNS can be used to improve the localization resolution attained by previous work in \cite{doa_2_mic_app1, doa_3_mic_app1,doa_3_mic_app2}.\par

\begin{figure*}[htb!]
  \centering
  \begin{tabular}{ c c c }
    \includegraphics[width=0.3\textwidth]{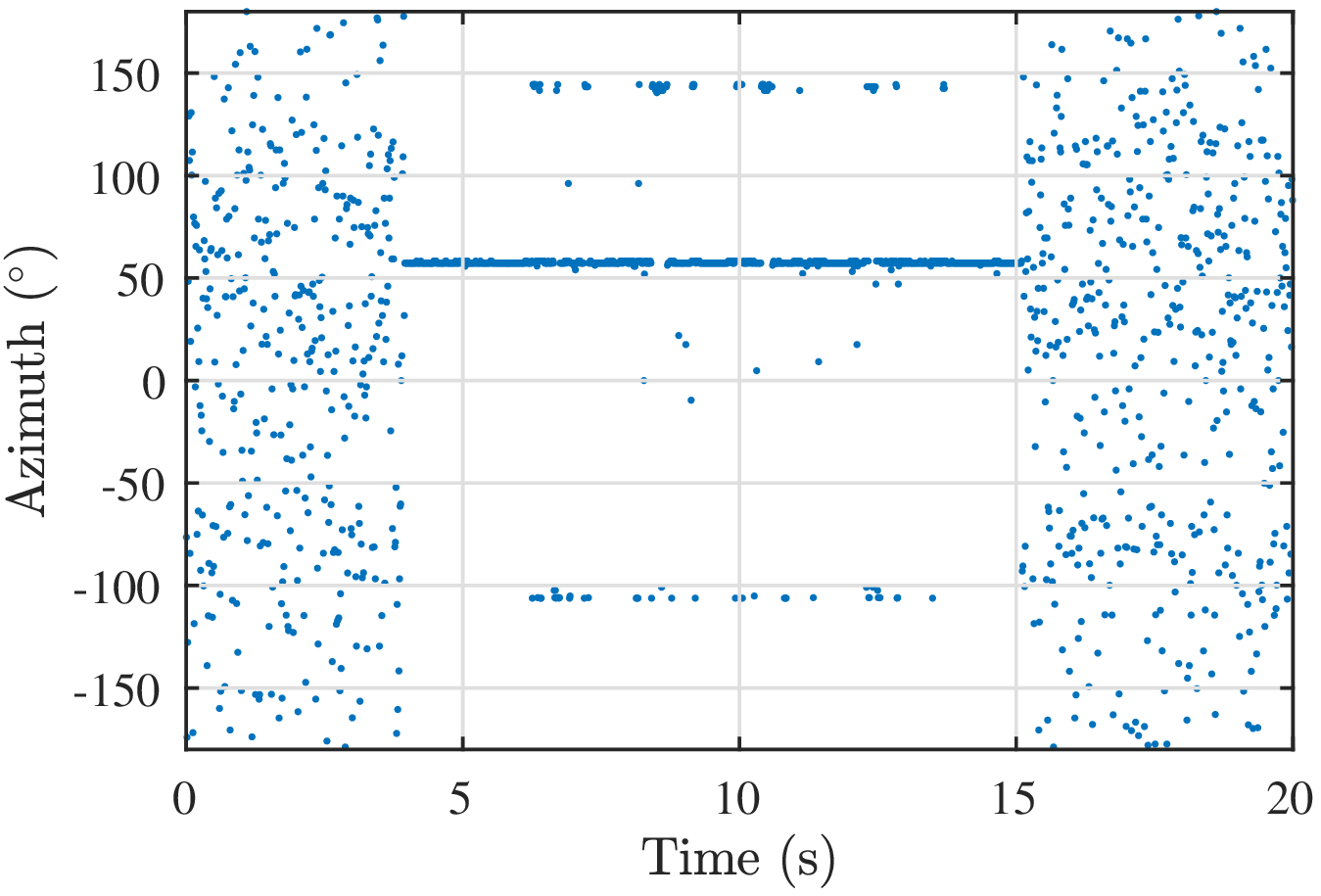}  & \includegraphics[width=0.3\textwidth]{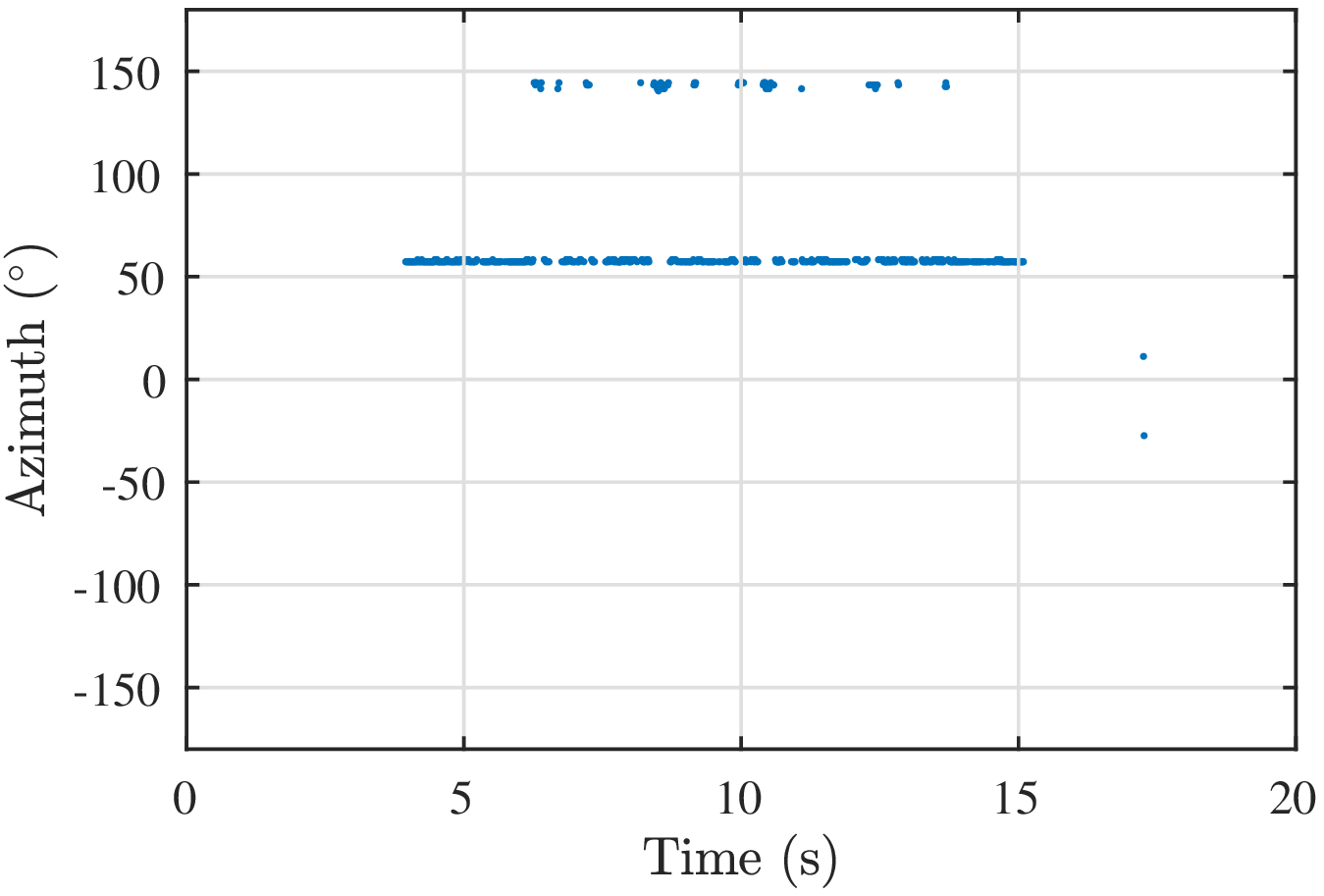} & \includegraphics[width=0.3\textwidth]{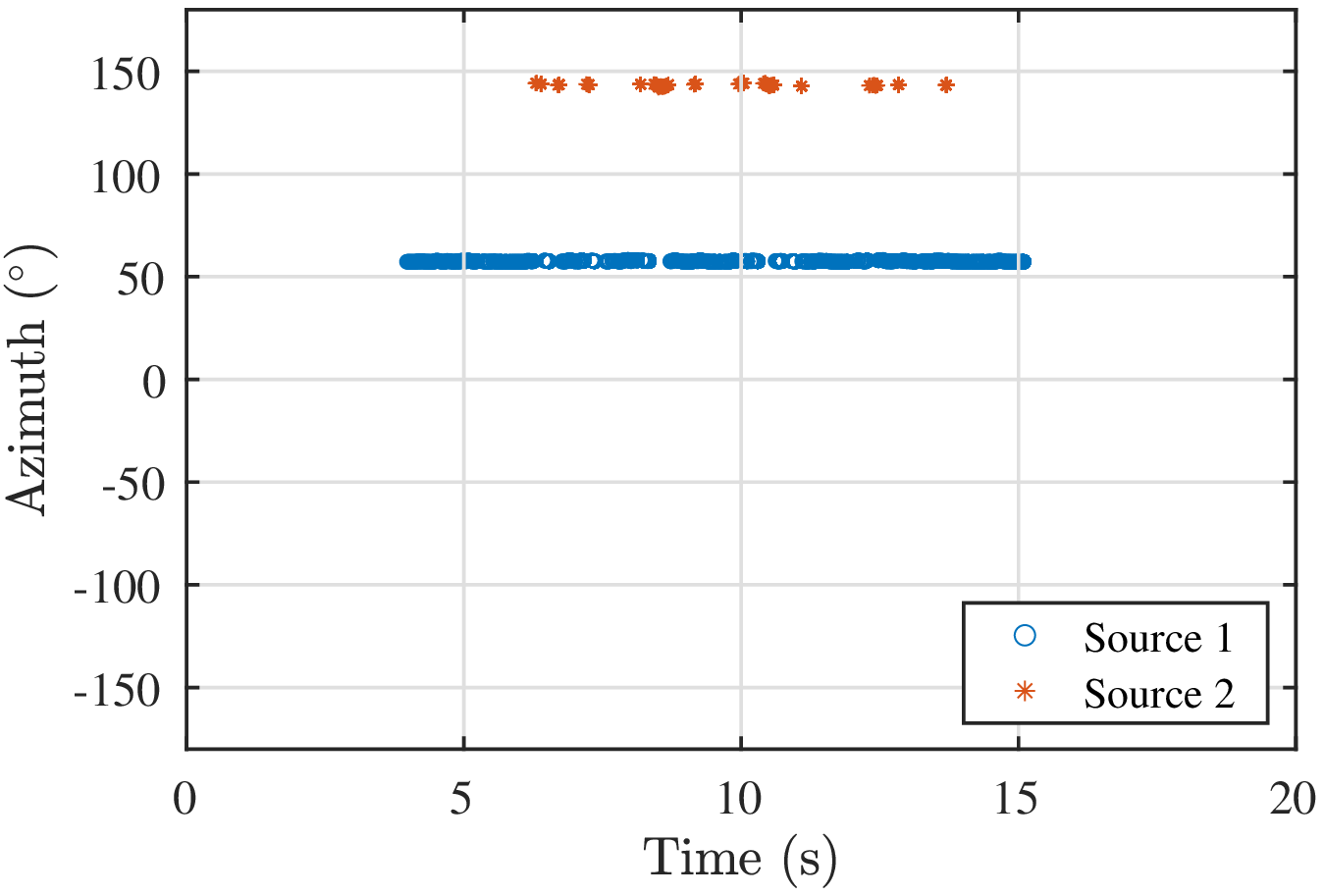} \\ \small (a) & \small (b) & \small (c)
  \end{tabular}
  \caption{Experiment 2. Processing stages in detection and localization of two overlapping sources. Elevation omitted for conciseness. Music and speech start at 4 and 6 seconds and end at 14 and 15 seconds, respectively. (a) Raw DOA estimates. (b) Filtered estimates. (c) Filtered and clustered estimates.}
  \label{fig:filter_cluster_doa}
\end{figure*}

In the second experiment, we test the capability of the proposed method to detect and localize two overlapping sound sources. One source was a speaker playing music, which was placed at two-meter distance, $55^\circ$ azimuth, and $0^\circ$ elevation. The other source was a human speaker talking, which stood at half a meter distance, $145^\circ$ azimuth and $40^\circ$ elevation. Results are shown in Fig. \ref{fig:filter_cluster_doa}. The frame-level localization estimates are mostly accurate except for a ``ghost" source at $100^\circ$, which we attribute to smoothing in the cross-correlation function caused by multiple dominant sources in a frame. However, filtering removes most outliers. Finally, RFEFC detects the two sources and, as a result of having $N_s > 1$, adds an additional layer of filtering by dismissing remaining outliers. We note that, unlike \cite{doa_2_mic_app1, doa_3_mic_app1,doa_3_mic_app2}, no voice activity detector is used, allowing detection and localization of sources other than speech.\par

\begin{figure}[htb!]
    \centering
    \includegraphics[trim={0 0 0 0},clip, width=0.32\textwidth]{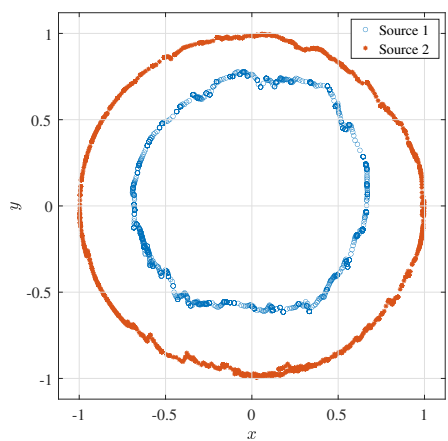}
    \caption{Experiment 3. Clustered location estimates on an orthographic projection centered at the pole of a unit radius hemisphere.}
    \label{fig:tracking_doa}
\end{figure}

As in RTEFC, the use of exponential filtering in RFEFC suggests that the proposed method may in principle be capable of tracking moving sources producing continuous sound. To verify this claim, in the third and last experiment, we test the ability of the proposed method to track two overlapping sources. The same two-source localization scenario is considered as in the second experiment with the difference that the smartphone is placed on a rotating turntable making a full revolution in 20 seconds. Hence, we expect source positions to form two concentric circles on a hemisphere, with each circle being defined according to the elevation of a respective source. Results in Fig. \ref{fig:tracking_doa} suggest that the proposed method indeed exhibits excellent tracking capabilities.\par

\section{Conclusion}
A method for accurate 2D-DOA estimation using a nonlinear 3-microphone array was proposed. The problem is modeled as localization of one or more acoustic sources on a unit-radius hemisphere above the array. For best practical results, the derived CF is replaced with NNS applied to a spherical Fibonacci lattice containing TDOA to 2D-DOA mappings learned in the field. Filtering and clustering post-processors were also introduced to reject unreliable measurements and allow more robust detection and localization of multiple sound sources. When evaluated in the field, the proposed method displayed remarkable 2D-DOA estimation accuracy and tracking capabilities of two overlapping sources.\par


\bibliographystyle{IEEEtran}
\bibliography{mybib}

\end{document}